# Compression-Based Investigation of the Dynamical Properties of Cellular Automata and Other Systems

**Hector Zenil**

*Laboratoire d'Informatique Fondamentale de Lille (USTL/CNRS)*
*and*
*Institut d'Histoire et de Philosophie des Sciences et des Techniques*
*(CNRS/Paris 1/ENS Ulm)*
*hector.zenil-chavez@malix.univ-paris1.fr*

A method for studying the qualitative dynamical properties of abstract computing machines based on the approximation of their program-size complexity using a general lossless compression algorithm is presented. It is shown that the compression-based approach classifies cellular automata into clusters according to their heuristic behavior. These clusters show a correspondence with Wolfram's main classes of cellular automata behavior. A Gray code-based numbering scheme is developed for distinguishing initial conditions. A compression-based method to estimate a characteristic exponent for detecting phase transitions and measuring the resiliency or sensitivity of a system to its initial conditions is also proposed. A conjecture regarding the capability of a system to reach computational universality related to the values of this phase transition coefficient is formulated. These ideas constitute a compression-based framework for investigating the dynamical properties of cellular automata and other systems.

## 1. Introduction

Previous investigations of the dynamical properties of cellular automata have involved compression in one form or another. In this paper, we take the direct approach, experimentally studying the relationship between properties of dynamics and their compression.

Cellular automata were first introduced by J. von Neumann [1] as a mathematical model for biological self-replication phenomena. They have since played a basic role in understanding and explaining various complex physical, social, chemical, and biological phenomena. Using extensive computer simulation S. Wolfram [2] classified cellular automata into four classes according to the qualitative behavior of their evolution. This classification has been further investigated and verified by G. Braga et al. [3], followed by more detailed verifications and investigations of classes 1, 2, and 3 in [4, 5].

Other formal approaches to the problem of classifying cellular automata have also been attempted, with some success. Of these,

some are based on the structure of attractors or other topological classifications [6, 7], others use probabilistic approaches [8] or involve looking at whether a cellular automaton falls into some chaotic attractor or an undecidable class [9-11], while yet others use the idea of approaching the algorithmic or program-size complexity ($K$) of the rule table of a cellular automaton [12]. It has also been shown that rules that in certain conditions belong to one class may belong to another when starting from a different set up. This has been the case of rule 40, simple and therefore in class 1 when starting from a 0-finite configuration but chaotic [13] when starting from certain random configurations.

Compression-based mathematical characterizations and techniques for classifying and clustering have been suggested and successfully developed in areas as diverse as languages, literature, genomics, music, and astronomy. A good introduction can be found in [14]. Compression is a powerful tool for pattern recognition and has often been used for classification and clustering. Lempel-Ziv (LZ)-like data compressors have been proven to be universally optimal and are therefore good candidates as approximators of the program-size complexity of strings.

The program-size complexity [15] $K_u(s)$ of a string $s$ with respect to a universal Turing machine $U$ is defined as the binary length of the shortest program $p$ that produces as output the string $s$. Or, as a *Mathematica* expression:

$$K_u(s) = \{\min(\text{Length}[p]), \ U(p) = s\}$$

However, a drawback of $K$ is that it is an uncomputable function. In general, the only way to approach $K$ is by compressibility methods. Essentially, the program-size complexity of a string is the ultimate compressed version of that string.

As an attempt to capture and systematically study the behavior of abstract machines, our experimental approach consists of calculating the program-size complexity of the output of the evolution of a cellular automaton. This is done following methods of extended computation, enumerating, and exhaustively running the systems as suggested in [2].

A cellular automaton is a collection of cells on a grid of specified shape that evolves through a number of discrete time steps according to a set of rules based on the states of neighboring cells. The rules are applied iteratively for as many time steps as desired. The number of colors (or distinct states) $k$ of a cellular automaton is a non-negative integer. In addition to the grid on which a cellular automaton lives and the colors its cells may assume, the neighborhood over which cells affect one another must also be specified. The simplest choice is a nearest-neighbor rule, in which only cells directly adjacent to a given cell may be affected at each time step. The simplest type of cellular automaton is then a binary, nearest-neighbor, one-dimensional automaton (called *elementary* by Wolfram). There are 256 such automata, each of which can be indexed by a unique binary number whose decimal representation is known as the rule for the particular automaton.

Regardless of the apparent simplicity of their formal description, cellular automata are capable of displaying a wide range of interesting and different dynamical properties as thoroughly investigated by Wolfram in [2]. The problem of classification is a central topic in cellular automata theory.

Wolfram identifies and classifies cellular automata (and other discrete systems) as displaying these four different classes of behavior.

1. A fixed, homogeneous state is eventually reached (e.g., rules 0, 8, 136).

2. A pattern consisting of separated periodic regions is produced (e.g., rules 4, 37, 56, 73).

3. A chaotic, aperiodic pattern is produced (e.g., rules 18, 45, 146).

4. Complex, localized structures are generated (e.g., rule 110).

## 2. Compression-Based Classification

The method consists of compressing the evolution of a cellular automaton up to a certain number of steps. The *Mathematica* function `Compress` [16] gives a compressed representation of an expression as a string. It uses a C language implementation of a "deflate" compliant compressor and decompressor available within the zlib package. The deflate lossless compression algorithm, independent of CPU type, operating system, file system, and character set compresses data using a combination of the LZ algorithm and Huffman coding [17-19], with efficiency comparable to the best currently available general-purpose compression methods as described in RFC 1951 [20] called the Lempel-Ziv-Welch (LZW) algorithm. The same algorithm is the basis of the widely used gzip data compression software. Data compression is generally achieved through two steps:

- The matching and replacement of duplicate strings with pointers.
- Replacing symbols with new, weighted symbols based on frequency of use.

### 2.1 Compression-Based Classification of Elementary Cellular Automata from Simplest Initial Conditions

The difference in length between the compressed and uncompressed forms of the output of a cellular automaton is a good approximation of its program-size complexity. In most cases, the length of the compressed form levels off, indicating that the cellular automaton output is repetitive and can easily be described. However, in cases like rules 30, 45, 110, or 73 the length of the compressed form grows rapidly, corresponding to the apparent randomness and lack of structure in the display.

### 2.1.1 Classification Parameters

There are two main parameters that play a role when classifying cellular automata: the initial configuration and the number of steps. Classifying cellular automata can begin by starting all with a single black cell. Some of them, such as rule 30, will immediately show their full richness in dynamical terms, while others might produce very different behavior when starting with another initial configuration. Both types might produce different classifications. We first explore the case of starting with a single black cell and then proceed to consider the other case for detecting phase transitions.

An illustration of the evolution of rules 95, 82, 50, and 30 is shown in Figure 1, together with the compressed and uncompressed lengths they produce, each starting from a single black cell moving through time (number of steps).

As shown in Figure 1, the compressed lengths of simple cellular automata do not approach the uncompressed lengths and stay flat or grow linearly, while the length of the compressed form approaches the length of the uncompressed form for rules such as 30.

Cellular automata can be classified by sorting their compressed lengths as an approximation to their program-size complexity. In Figure 2, $c$ is the compressed length of the evolution of the cellular automaton up to the first 200 steps (although the pictures only show the first 60).

Early in 2007 I wrote a program using *Mathematica* to illustrate the basic idea of the compressibility method for classifying cellular automata. The program (called a Demonstration) was submitted to the Wolfram Demonstrations Project and published under the title "Cellular Automaton Compressibility" [21]. Later in 2007, inspired by this Demonstration and under my mentorship, Joe Bolte from Wolfram Research, Inc. developed a project under the title "Automatic Ranking and Sorting of Cellular Automaton Complexity" at the NKS Summer School held at the University of Vermont (for more information see http://www.wolframscience.com/summerschool/2007/participants/bolte.html). In 2009, also under my mentorship, Chiara Basile from the University of Bologna would further develop the project at the NKS Summer School held at the ISTI-CNR in Pisa, Italy, under the title "Exploring CA Rule Spaces by Means of Data Compression." The project was enriched by Basile's own prior research, particularly on feeding the compression algorithm with sequences following different directions (rows, columns, and space-filling curves), thus helping speed up the pattern detection process (for more information see http://www.wolframscience.com/summerschool/2009/participants/basile.html).

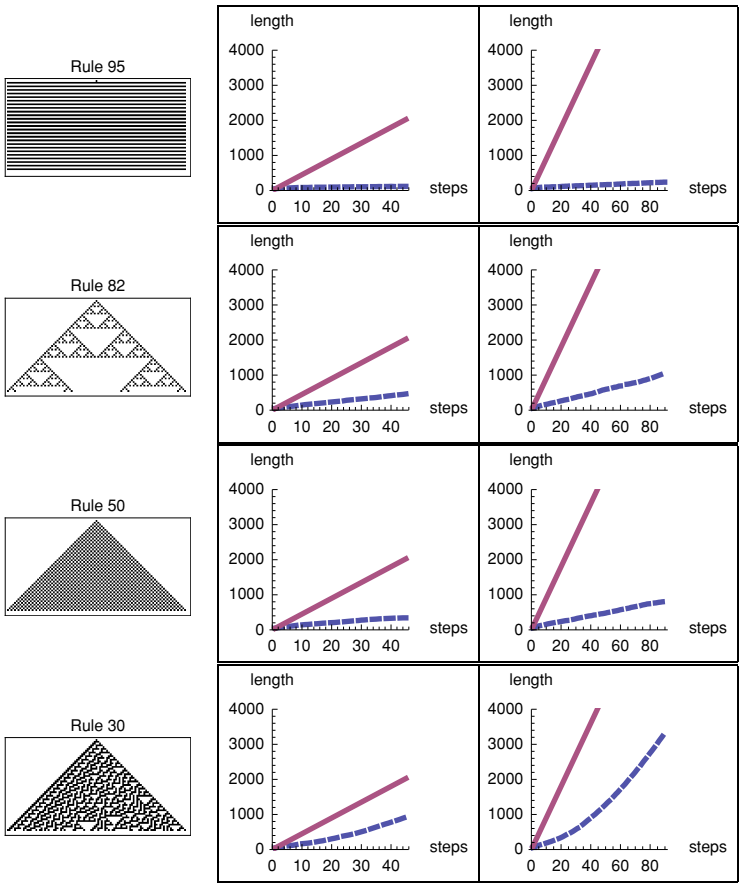

**Figure 1**. Evolution of rules 95, 82, 50, and 30 together with the compressed (dashed line) and uncompressed (solid line) lengths.

**Figure 2**. Complete compression-based classification of the elementary cellular automata (*continues*).

| rule 104 | rule 96 | rule 95 | rule 87 | rule 72 | rule 64 |
|---|---|---|---|---|---|
| 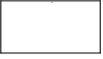 | 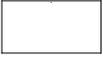 | 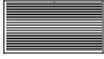 | 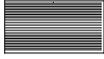 | 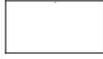 | 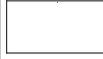 |
| $c = 1026$ | $c = 1026$ | $c = 1026$ | $c = 1026$ | $c = 1026$ | $c = 1026$ |
| rule 63 | rule 55 | rule 40 | rule 32 | rule 31 | rule 23 |
| 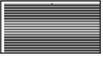 | 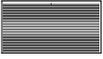 | 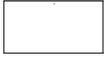 | 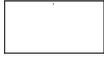 | 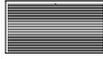 | 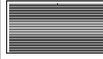 |
| $c = 1026$ | $c = 1026$ | $c = 1026$ | $c = 1026$ | $c = 1026$ | $c = 1026$ |
| rule 8 | rule 0 | rule 253 | rule 251 | rule 239 | rule 21 |
| 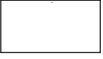 | 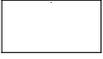 | 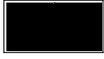 | 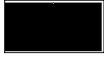 | 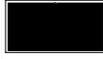 | 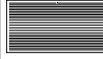 |
| $c = 1026$ | $c = 1026$ | $c = 1030$ | $c = 1030$ | $c = 1030$ | $c = 1030$ |
| rule 19 | rule 7 | rule 237 | rule 249 | rule 235 | rule 233 |
| 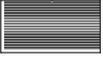 | 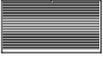 | 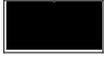 | 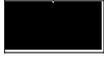 | 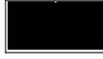 | 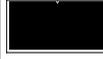 |
| $c = 1030$ | $c = 1030$ | $c = 1034$ | $c = 1038$ | $c = 1038$ | $c = 1046$ |
| rule 221 | rule 205 | rule 236 | rule 228 | rule 219 | rule 217 |
| 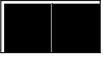 | 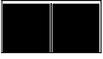 | 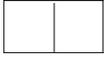 | 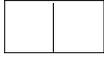 | 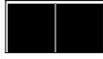 | 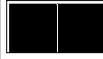 |
| $c = 1490$ | $c = 1490$ | $c = 1562$ | $c = 1562$ | $c = 1562$ | $c = 1562$ |
| rule 207 | rule 204 | rule 203 | rule 201 | rule 196 | rule 172 |
| 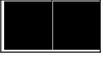 | 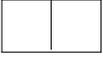 | 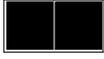 | 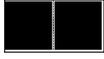 | 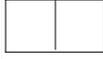 | 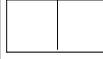 |
| $c = 1562$ | $c = 1562$ | $c = 1562$ | $c = 1562$ | $c = 1562$ | $c = 1562$ |
| rule 164 | rule 140 | rule 132 | rule 108 | rule 100 | rule 76 |
| 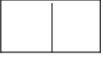 | 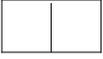 | 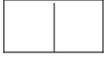 | 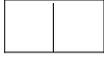 | 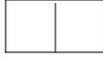 | 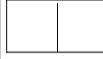 |
| $c = 1562$ | $c = 1562$ | $c = 1562$ | $c = 1562$ | $c = 1562$ | $c = 1562$ |
| rule 68 | rule 51 | rule 44 | rule 36 | rule 12 | rule 4 |
| 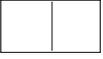 | 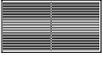 | 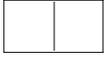 | 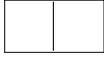 | 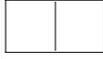 | 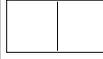 |
| $c = 1562$ | $c = 1562$ | $c = 1562$ | $c = 1562$ | $c = 1562$ | $c = 1562$ |
| rule 29 | rule 71 | rule 123 | rule 33 | rule 5 | rule 1 |
| 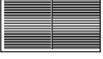 | 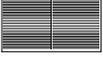 | 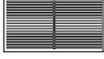 | 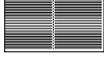 | 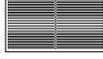 | 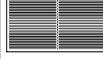 |
| $c = 1566$ | $c = 1582$ | $c = 1606$ | $c = 1606$ | $c = 1606$ | $c = 1606$ |
| rule 91 | rule 37 | rule 83 | rule 53 | rule 234 | rule 226 |
| 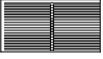 | 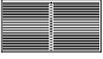 | 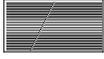 | 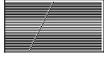 | 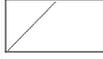 | 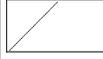 |
| $c = 1610$ | $c = 1614$ | $c = 1786$ | $c = 1786$ | $c = 1790$ | $c = 1790$ |
| rule 202 | rule 194 | rule 189 | rule 187 | rule 175 | rule 170 |
| 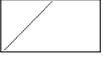 | 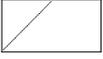 | 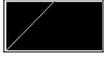 | 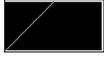 | 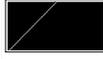 | 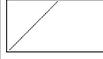 |
| $c = 1790$ | $c = 1790$ | $c = 1790$ | $c = 1790$ | $c = 1790$ | $c = 1790$ |
| rule 162 | rule 138 | rule 130 | rule 106 | rule 98 | rule 85 |
| 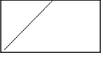 | 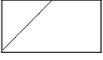 | 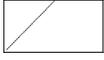 | 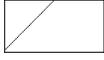 | 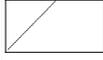 | 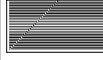 |
| $c = 1790$ | $c = 1790$ | $c = 1790$ | $c = 1790$ | $c = 1790$ | $c = 1790$ |

**Figure 2.** (*continued*)

| rule 74 | rule 66 | rule 42 | rule 39 | rule 34 | rule 27 |
|---|---|---|---|---|---|
| 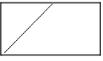 | 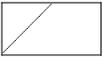 | 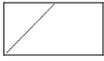 | 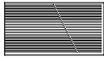 | 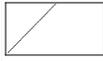 | 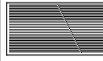 |
| *c = 1790* | *c = 1790* | *c = 1790* | *c = 1790* | *c = 1790* | *c = 1790* |
| rule 10 | rule 2 | rule 248 | rule 245 | rule 243 | rule 240 |
| 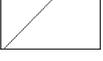 | 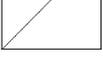 | 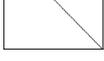 | 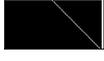 | 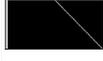 | 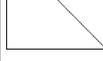 |
| *c = 1790* | *c = 1790* | *c = 1794* | *c = 1794* | *c = 1794* | *c = 1794* |
| rule 231 | rule 216 | rule 208 | rule 184 | rule 176 | rule 174 |
| 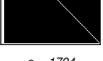 | 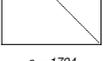 | 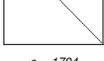 | 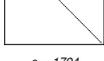 | 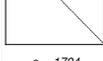 | 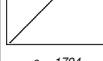 |
| *c = 1794* | *c = 1794* | *c = 1794* | *c = 1794* | *c = 1794* | *c = 1794* |
| rule 173 | rule 152 | rule 144 | rule 143 | rule 142 | rule 120 |
| 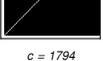 | 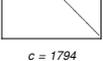 | 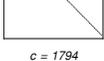 | 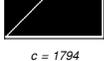 | 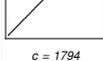 | 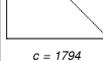 |
| *c = 1794* | *c = 1794* | *c = 1794* | *c = 1794* | *c = 1794* | *c = 1794* |
| rule 117 | rule 113 | rule 112 | rule 88 | rule 81 | rule 80 |
| 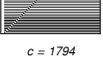 | 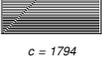 | 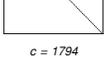 | 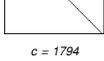 | 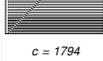 | 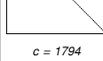 |
| *c = 1794* | *c = 1794* | *c = 1794* | *c = 1794* | *c = 1794* | *c = 1794* |
| rule 56 | rule 48 | rule 46 | rule 24 | rule 16 | rule 15 |
| 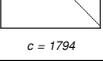 | 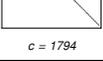 | 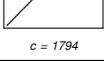 | 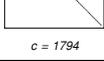 | 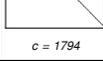 | 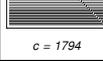 |
| *c = 1794* | *c = 1794* | *c = 1794* | *c = 1794* | *c = 1794* | *c = 1794* |
| rule 14 | rule 244 | rule 241 | rule 229 | rule 227 | rule 213 |
| 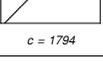 | 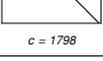 | 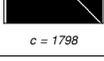 | 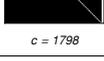 | 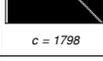 | 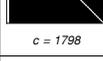 |
| *c = 1794* | *c = 1798* | *c = 1798* | *c = 1798* | *c = 1798* | *c = 1798* |
| rule 212 | rule 209 | rule 185 | rule 171 | rule 139 | rule 116 |
| 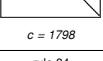 | 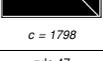 | 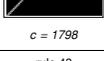 | 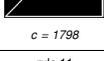 | 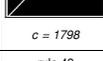 | 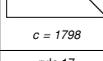 |
| *c = 1798* | *c = 1798* | *c = 1798* | *c = 1798* | *c = 1798* | *c = 1798* |
| rule 84 | rule 47 | rule 43 | rule 11 | rule 49 | rule 17 |
| 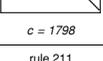 | 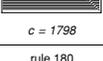 | 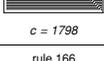 | 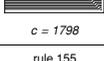 | 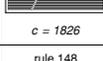 | 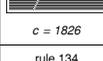 |
| *c = 1798* | *c = 1798* | *c = 1798* | *c = 1798* | *c = 1826* | *c = 1826* |
| rule 211 | rule 180 | rule 166 | rule 155 | rule 148 | rule 134 |
| 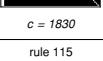 | 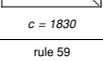 | 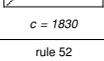 | 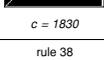 | 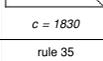 | 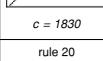 |
| *c = 1830* | *c = 1830* | *c = 1830* | *c = 1830* | *c = 1830* | *c = 1830* |
| rule 115 | rule 59 | rule 52 | rule 38 | rule 35 | rule 20 |
| 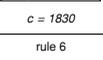 | 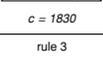 | 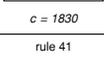 | 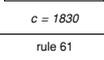 | 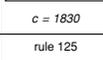 | 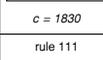 |
| *c = 1830* | *c = 1830* | *c = 1830* | *c = 1830* | *c = 1830* | *c = 1830* |
| rule 6 | rule 3 | rule 41 | rule 61 | rule 125 | rule 111 |
| 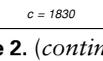 | 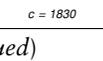 | 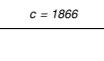 | 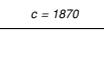 | 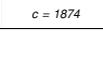 | 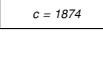 |
| *c = 1830* | *c = 1830* | *c = 1866* | *c = 1870* | *c = 1874* | *c = 1874* |

**Figure 2.** (*continued*)

| rule 103 | rule 67 | rule 65 | rule 25 | rule 9 | rule 97 |
|---|---|---|---|---|---|
| 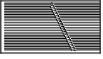 | 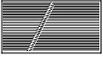 | 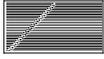 | 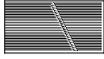 | 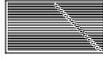 | 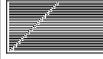 |
| *c = 1874* | *c = 1874* | *c = 1874* | *c = 1874* | *c = 1874* | *c = 1878* |
| rule 107 | rule 121 | rule 141 | rule 78 | rule 252 | rule 220 |
| 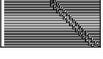 | 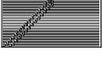 | 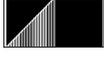 | 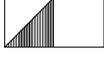 | 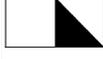 | 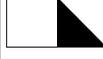 |
| *c = 1930* | *c = 1938* | *c = 2370* | *c = 2386* | *c = 2434* | *c = 2434* |
| rule 79 | rule 13 | rule 238 | rule 206 | rule 69 | rule 93 |
| 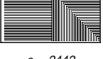 | 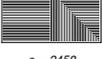 | 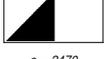 | 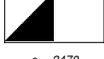 | 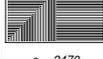 | 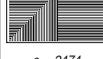 |
| *c = 2442* | *c = 2458* | *c = 2470* | *c = 2470* | *c = 2470* | *c = 2474* |
| rule 254 | rule 222 | rule 157 | rule 198 | rule 70 | rule 163 |
| 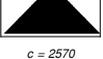 | 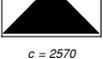 | 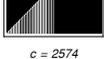 | 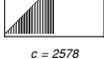 | 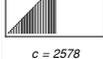 | 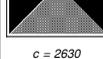 |
| *c = 2570* | *c = 2570* | *c = 2574* | *c = 2578* | *c = 2578* | *c = 2630* |
| rule 199 | rule 250 | rule 242 | rule 186 | rule 179 | rule 178 |
| 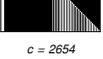 | 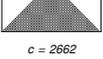 | 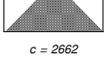 | 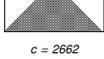 | 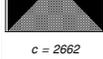 | 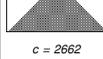 |
| *c = 2654* | *c = 2662* | *c = 2662* | *c = 2662* | *c = 2662* | *c = 2662* |
| rule 156 | rule 122 | rule 114 | rule 77 | rule 58 | rule 50 |
| 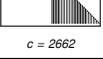 | 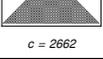 | 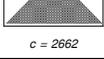 | 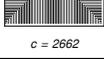 | 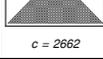 | 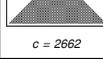 |
| *c = 2662* | *c = 2662* | *c = 2662* | *c = 2662* | *c = 2662* | *c = 2662* |
| rule 28 | rule 188 | rule 230 | rule 225 | rule 197 | rule 246 |
| 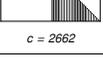 | 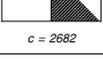 | 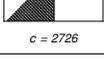 | 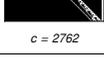 | 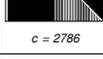 | 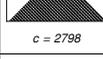 |
| *c = 2662* | *c = 2682* | *c = 2726* | *c = 2762* | *c = 2786* | *c = 2798* |
| rule 92 | rule 190 | rule 169 | rule 147 | rule 54 | rule 158 |
| 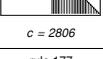 | 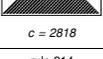 | 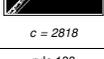 | 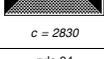 | 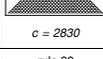 | 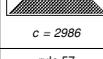 |
| *c = 2806* | *c = 2818* | *c = 2818* | *c = 2830* | *c = 2830* | *c = 2986* |
| rule 177 | rule 214 | rule 133 | rule 94 | rule 99 | rule 57 |
| 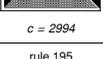 | 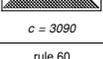 | 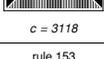 | 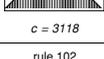 | 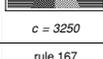 | 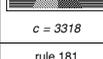 |
| *c = 2994* | *c = 3090* | *c = 3118* | *c = 3118* | *c = 3250* | *c = 3318* |
| rule 195 | rule 60 | rule 153 | rule 102 | rule 167 | rule 181 |
| 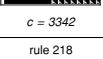 | 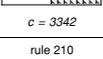 | 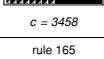 | 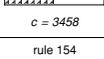 | 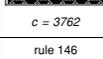 | 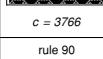 |
| *c = 3342* | *c = 3342* | *c = 3458* | *c = 3458* | *c = 3762* | *c = 3766* |
| rule 218 | rule 210 | rule 165 | rule 154 | rule 146 | rule 90 |
| 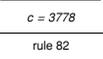 | 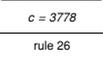 | 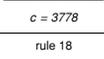 | 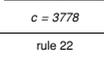 | 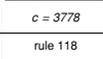 | 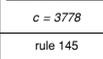 |
| *c = 3778* | *c = 3778* | *c = 3778* | *c = 3778* | *c = 3778* | *c = 3778* |
| rule 82 | rule 26 | rule 18 | rule 22 | rule 118 | rule 145 |
| 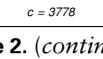 | 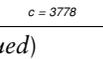 | 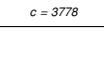 | 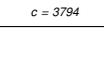 | 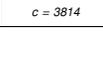 | 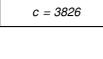 |
| *c = 3778* | *c = 3778* | *c = 3778* | *c = 3794* | *c = 3814* | *c = 3826* |

**Figure 2.** (*continued*)

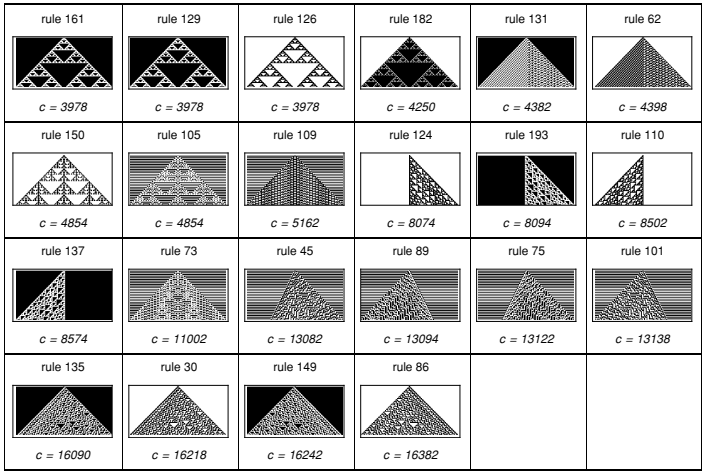

**Figure 2.** (*continued*)

## 3. Compression-Based Clustering

### 3.1 2-Clusters Plot

By finding neighboring clusters of compressed lengths cellular automata can be grouped by their program-size complexity. Treating pairs of elements as being less similar when their distances are larger using an Euclidean distance function, a 2-clusters plot was able to separate cellular automata that clearly fall into Wolfram's classes 3 and 4 from the rest, dividing complex and random-looking cellular automata from trivial and nested ones as shown in Figures 3 through 5.

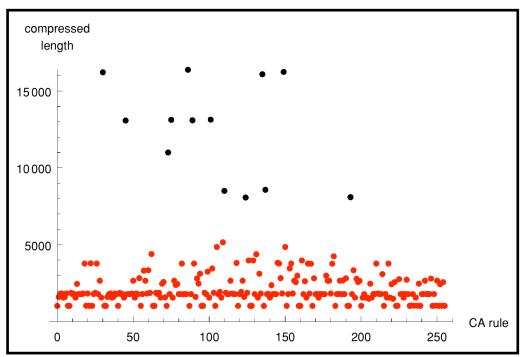

**Figure 3.** Partitioning elementary cellular automata into clusters by compression length.

| Classes 1 and 2 | | | | | | | | | | | | | | |
|---|---|---|---|---|---|---|---|---|---|---|---|---|---|---|
| 0 | 1 | 2 | 3 | 4 | 5 | 6 | 7 | 8 | 9 | 10 | 11 | 12 | 13 | 14 |
| 15 | 16 | 17 | 18 | 19 | 20 | 21 | 22 | 23 | 24 | 25 | 26 | 27 | 28 | 29 |
| 31 | 32 | 33 | 34 | 35 | 36 | 37 | 38 | 39 | 40 | 41 | 42 | 43 | 44 | 46 |
| 47 | 48 | 49 | 50 | 51 | 52 | 53 | 54 | 55 | 56 | 57 | 58 | 59 | 60 | 61 |
| 62 | 63 | 64 | 65 | 66 | 67 | 68 | 69 | 70 | 71 | 72 | 74 | 76 | 77 | 78 |
| 79 | 80 | 81 | 82 | 83 | 84 | 85 | 87 | 88 | 90 | 91 | 92 | 93 | 94 | 95 |
| 96 | 97 | 98 | 99 | 100 | 102 | 103 | 104 | 105 | 106 | 107 | 108 | 109 | 111 | 112 |
| 113 | 114 | 115 | 116 | 117 | 118 | 119 | 120 | 121 | 122 | 123 | 125 | 126 | 127 | 128 |
| 129 | 130 | 131 | 132 | 133 | 134 | 136 | 138 | 139 | 140 | 141 | 142 | 143 | 144 | 145 |
| 146 | 147 | 148 | 150 | 151 | 152 | 153 | 154 | 155 | 156 | 157 | 158 | 159 | 160 | 161 |
| 162 | 163 | 164 | 165 | 166 | 167 | 168 | 169 | 170 | 171 | 172 | 173 | 174 | 175 | 176 |
| 177 | 178 | 179 | 180 | 181 | 182 | 183 | 184 | 185 | 186 | 187 | 188 | 189 | 190 | 191 |
| 192 | 194 | 195 | 196 | 197 | 198 | 199 | 200 | 201 | 202 | 203 | 204 | 205 | 206 | 207 |
| 208 | 209 | 210 | 211 | 212 | 213 | 214 | 215 | 216 | 217 | 218 | 219 | 220 | 221 | 222 |
| 223 | 224 | 225 | 226 | 227 | 228 | 229 | 230 | 231 | 232 | 233 | 234 | 235 | 236 | 237 |
| 238 | 239 | 240 | 241 | 242 | 243 | 244 | 245 | 246 | 247 | 248 | 249 | 250 | 251 | 252 |
| 253 | 254 | 255 | | | | | | | | | | | | |

| Classes 3 and 4 | | |
|---|---|---|
| 30 | 45 | 73 |
| 75 | 86 | 89 |
| 101 | 110 | 124 |
| 135 | 137 | 149 |
| 193 | | |

**Figure 4**. Elementary cellular automata clusters by compressibility.

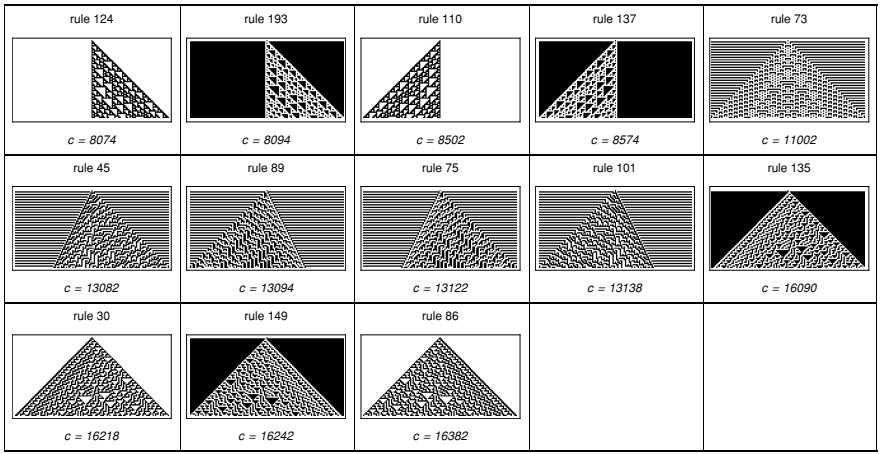

**Figure 5**. Elementary cellular automata classes 3 and 4.

A second application of the clustering algorithm splits the original classes 3 and 4 into clusters linking automata by their qualitative properties as shown in Figure 6.

## 3.2 Compression-Based Classification of Larger Spaces of Cellular Automata and Other Abstract Machines
### 3.2.1 3-Color Nearest-Neighbor Cellular Automata

By following the same technique, we were able to identify 3-color nearest-neighbor cellular automata in classes 3 and 4 as shown in Figure 7.

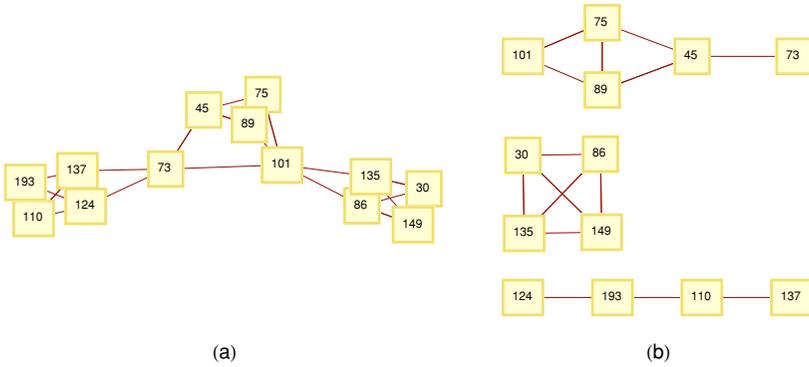

**Figure 6.** (a) Breaking class 3 and 4 clusters. (b) Splitting classes 3 and 4 by nearest compression lengths.

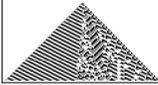

**Figure 7.** 3-color nearest-neighbor cellular automata compression-based search for classes 3 and 4 from a random sample.

### 3.2.2  2-State 3-Color Turing Machines

The exploration of Turing machines is considerably more difficult for three main reasons.

1. The spaces of Turing machines for the shortest states and colors are much larger than the shortest spaces of cellular automata.
2. Turing machines with nontrivial dynamical properties are very rare compared to the size of the space defined by the number of states and colors, and therefore larger samples are necessary.
3. Turing machines evolve much more slowly than cellular automata, so longer runtimes are also necessary.

The best compression-based results for identifying nontrivial Turing machines were obtained by applying the technique to the number of states a Turing machine is able to reach after a certain number of steps rather than to the output itself, unlike for cellular automata. The complexity of a Turing machine is deeply determined by the number of states the machine is capable of reaching from an initial configuration, and by looking at its complexity the technique distinguished the nontrivial machines from the most trivial. Figure 8 shows a sample of Turing machines found by applying the compression-based method.

## 4. Compression-Based Phase Transition Detection

A phase transition can be defined as a discontinuous change in the dynamical behavior of a system when a parameter associated with the system, such as its initial configuration, is varied.

It is conventional to distinguish two kinds of phase transitions, often called first- and higher-order. As described in [2], one feature of first-order transitions is that as soon as the transition is passed, the whole system always switches completely from one state to the other. However, higher-order transitions are gradual or recurrent. On one side of the transition a system is typically completely ordered or disordered. But when the transition is passed, the system does not change from then on to either one or another state. Instead, its order increases or decreases more or less gradually or recurrently as the parameter is varied. Typically the presence of order is signaled by the breaking of some kind of symmetry; for example, two rules explored in this section (rules 22 and 109) were found to be highly disturbed with recurrent phase transitions due to a symmetry breaking when starting with certain initial configurations.

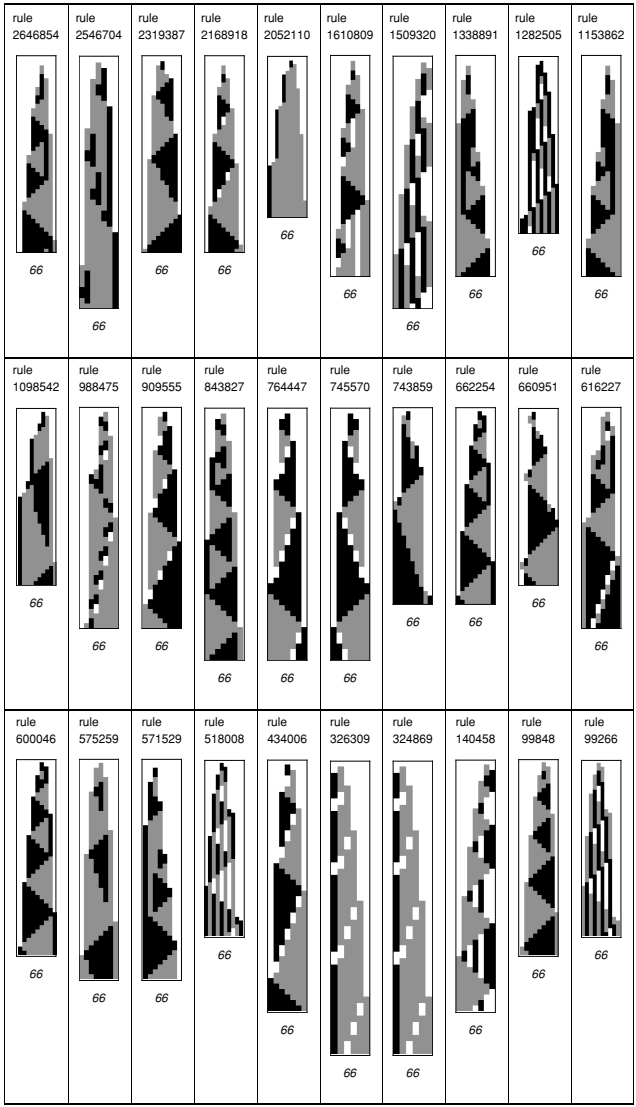

**Figure 8.** Compression-based search for nontrivial 2-state 3-color Turing machines.

## 4.1 Initial Configuration Numbering Scheme

Ideally, one should feed a system with a natural sequence of initial configurations of gradually increasing complexity. Doing so assures that qualitative changes in the evolution of the system are not attributable to discontinuities in its set of initial conditions.

The reflected binary code, also known as the "Gros-Gray code" or simply the "Gray code" (after Louis Gros and Frank Gray), is a binary numeral system where two successive values differ by only one bit. To explore the qualitative behavior of a cellular automaton when starting from different initial configurations, the optimal method is to follow a Gros-Gray encoding enumeration in order to avoid any undesirable "jumps" attributable to the system's having been fed with discontinuous initial configurations. By following the Gros-Gray code, an optimal numbering scheme was devised so that two consecutive initial conditions differ only by the simplest change (one bit).

`GrosGrayCodeDerivate` and `GrosGrayCodeIntegrate` implement the methods described in [22].

```
GrosGrayCodeDerivate[n_Integer] :=
 Prepend[Mod[#[[1]] + #[[2]], 2] & /@
    Partition[#, 2, 1], #[[1]]] &@ IntegerDigits[n, 2]

GrosGrayCodeIntegrate[l_List] :=
 FromDigits[Mod[#, 2] & /@ Accumulate[l], 2]
```

The function `InitialConfiguration` implements the optimal numbering scheme of initial conditions for cellular automata based on the Gros-Gray code, minimizing the Damerau-Levenshtein distance.

**Figure 9**. First 11 elements of the Gros-Gray code.

`GrosGrayCodeIntegrate` is the reverse function of `GrosGray`‎`CodeDerivate`. It retrieves the element number of an element in Gros-Gray's code, that is, the composition of `GrosGrayCodeDerivate` and `GrosGrayCodeIntegrate` is the identity function.

```
Table[GrosGrayCodeIntegrate[
   GrosGrayCodeDerivate[n]], {n, 0, 10}]
```

{0, 1, 2, 3, 4, 5, 6, 7, 8, 9, 10}

The Damerau-Levenshtein distance between two vectors $u$ and $v$ gives the number of one-element deletions, insertions, substitutions, and transpositions required to transform $u$ into $v$. It can be verified that the distance between any two adjacent elements in the Gros-Gray code is always 1.

```
DamerauLevenshteinDistance[#[[1]], #[[2]]] & /@
 Partition[Table[GrosGrayCodeDerivate[n], {n, 0, 10}], 2, 1]
```

{1, 1, 1, 1, 1, 1, 1, 1, 1, 1}

The simplest, not completely trivial, initial configuration of a cellular automaton is the typical single black cell that can be denoted (as in *Mathematica*) by {{1}, 0}, meaning a single black cell (1) on a background of whites (0). Preserving an "empty" background leaves the region that must be varied consisting only of the nonwhite portion of the initial configuration. However, when surrounded with zeroes, initial configurations may be the same for cellular automata. For example, the initial configuration {0, 1, 0} is exactly the same as {1} because the cellular automaton background is already filled with zeroes. Therefore, valid different initial configurations for cellular automata should always be wrapped in 1s.

```
InitialCondition[n_Integer] := If[n === 0, {Last[#]}, #] &@
   Append[GrosGrayCodeDerivate[n], 1]
```

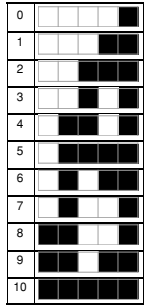

**Figure 10**. Sequence of the first 11 binary initial configurations for cellular automata based on the Gros-Gray code.

`InitialConditionNumber` is the reverse function for retrieving the number of an initial configuration given an initial configuration according to the numbering scheme devised herein.

```
InitialConditionNumber[l_List] := GrosGrayCodeIntegrate[Most[l]]
```

For example, the thirty-second initial condition is

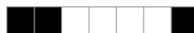

```
InitialConditionNumber[InitialCondition[32]]
```

32

An interesting example is the elementary cellular automaton rule 22, which behaves as a fractal in Wolfram's class 2 for some segment of initial configurations, followed by phase transitions of more disordered evolutions of the type of class 3.

## 4.2 Phase Transitions

Two one-dimensional elementary cellular automata that show discrete changes in behavior when the properties of their initial conditions are continuously changed are shown in Figures 11 and 12.

Data points are joined for clarity only. It can be seen that up to the initial configuration number 20 there are clear spikes at initial configurations 8, 14, 17, and 20 indicating four abrupt phase transitions.

For clarity, the background of the evolution of rule 109 in Figure 12 was cleaned up. Clear phase transitions are detected at initial configuration numbers 2, 3, 11, and 13, together with weaker behavior changes at initial configuration numbers 15, 16, and 20 that only occur on one side of the cellular automaton and therefore show spikes firing at half the length.

Comparison of the sequences of the compressed lengths of six different elementary cellular automata following the initial configuration numbering scheme up to the first $2^7 = 128$ initial configurations up to 150 steps each is shown in Figure 13.

The differences between the compressed versions provide information on the changes in behavior up to a given number of steps of a system starting from different initial conditions. The normalization divided by the number of steps provides the necessary stability to keep the increase of complexity on account of the increase of size due to longer runtimes out of the main equation. In other words, the program-size complexity accumulated due to longer runtimes is subtracted in time from the approximated program-size complexity of the system itself.

The method given can also be used to precompute the initial configurations of a cellular automaton space conducting the search for interesting behaviors and speeding up the study of qualitative dynamical properties. For example, interesting initial configurations to look at for rules 22 and 109 are those detected in Figure 13 showing clear phase transitions.

One open question is whether there are first-order phase transitions (when following a "natural" initial condition enumeration) in elementary cellular automata. Our method was only capable of detecting higher-order phase transitions up to the steps and initial conditions explored herein.

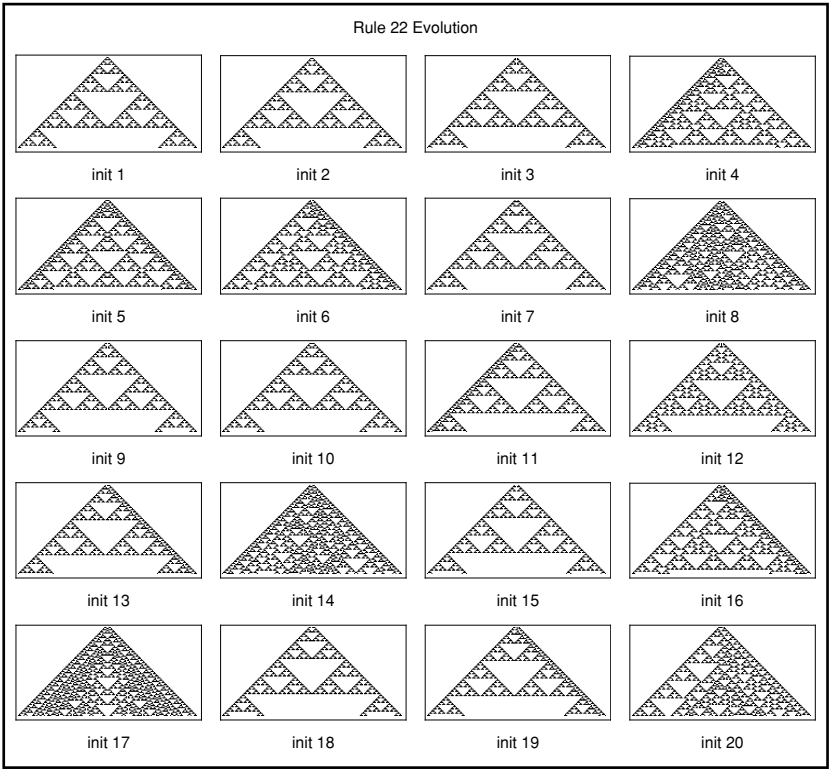

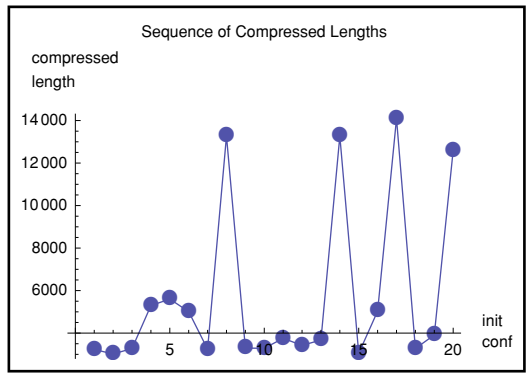

**Figure 11**. Rule 22 phase transition.

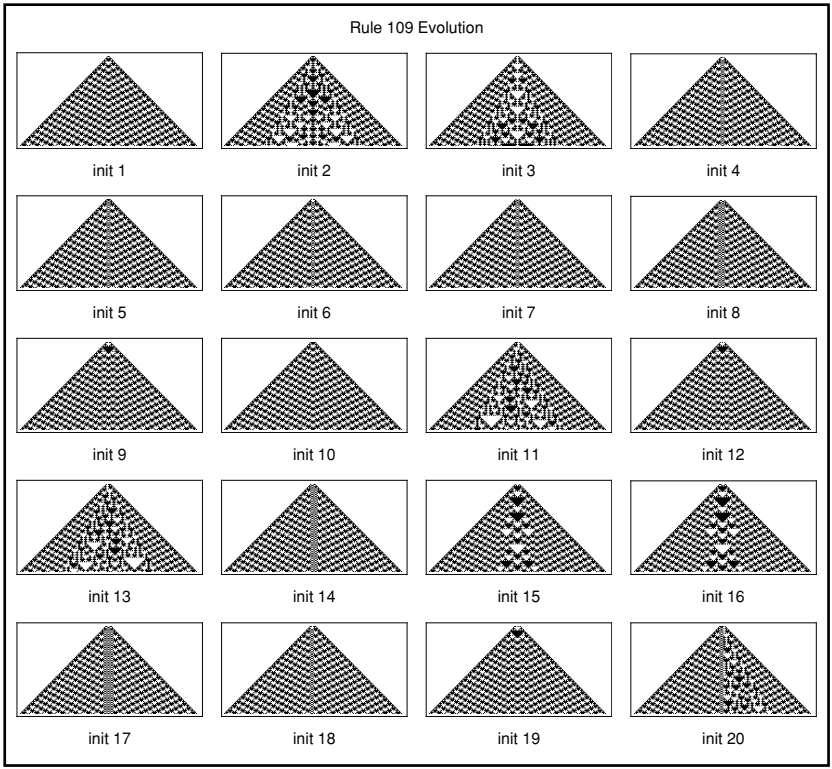

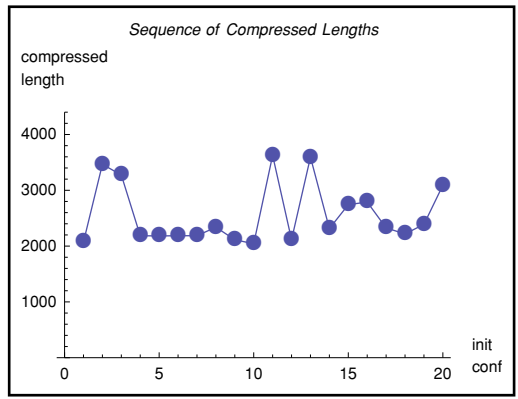

**Figure 12**. Rule 109 phase transition.

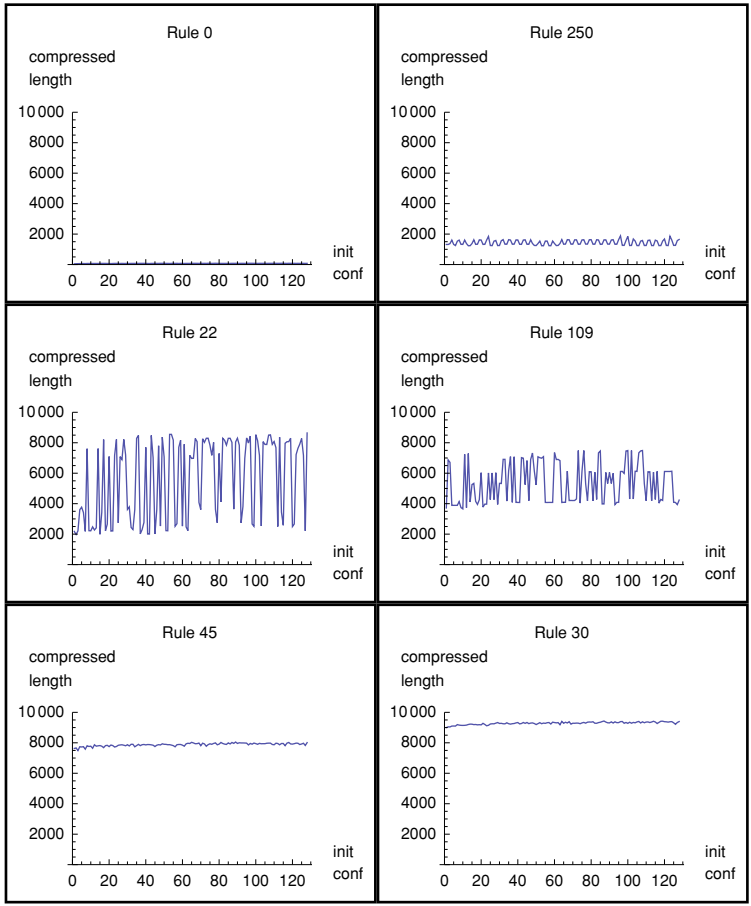

**Figure 13**. Sequence of compressed lengths for six elementary cellular automata.

## 5. Compression-Based Numerical Computation of a Characteristic Exponent

A fundamental property of chaotic behavior is the sensitivity to small changes in the initial conditions. Lyapunov characteristic exponents quantify this qualitative behavior by measuring the mean rate of divergence of initially neighboring trajectories. A characteristic exponent as a measure usually has the advantage of keeping systems with no significant phase transitions close to a constant value, while those with significant phase transitions are distinguished by a linear growth that characterizes instability in the system. Whether a system has a phase transition is an undecidable property of systems in general. However, the characteristic exponent is an effective method of calculation, even with no prior knowledge of the generating function of the system.

The technique described herein consists of comparing the mean of the divergence in time of the compressed lengths of the output of a system running over a sequence of small changes to the initial conditions over small intervals of time. The procedure yields a sequence of values normalized by the runtime and the derivative of the function that best fits the sequence. Just as with Wolfram's method described in [2] for the calculation of the Lyapunov exponents of a cellular automaton, the divergence in time is measured by the differences in space-time of the patterns produced by the system. But unlike the calculation of Lyapunov exponents, this will be done by measuring the distance between the compressed regions of the evolution of a cellular automaton when starting from different initial configurations. After normalization, we will be able to evaluate a stable characteristic exponent and therefore characterize its degree of sensitivity.

We want to examine the relative behavior of region evolutions when starting from adjacent initial configurations. Since the systems for which we are introducing this method are discrete, the regular parameter separating the initial conditions in a continuous system when calculating the Lyapunov exponent of a system can be replaced by the immediate successor of an initial condition following an optimal enumeration like the one described in Section 4 based on the Gros-Gray code.

Let the characteristic exponent $c_n^t$ be defined as the mean of the absolute values of the differences of the compressed lengths of the outputs of the system $M$ running over the initial segment of initial conditions $i_j$ with $j = \{1, \ldots, n\}$ following the numbering scheme devised earlier and running for $t$ steps, as follows:

$$c_n^t = \frac{|C(M_t(i_1)) - C(M_t(i_2))| + \cdots + |C(M_t(i_{n-1})) - C(M_t(i_n))|}{t(n-1)}$$

The division by $t$ acts as a normalization parameter in order to keep the runtime of the different values of $c_n^t$ as independent as possible among the systems. However, as already noted, normalization can also be achieved by dividing by the "volume" of the region (the space-time diagram) generated by a system (in the case of a one-dimensional cellular automaton the area, i.e., the number of affected cells—usually the characteristic cone). The mean of the absolute values can also be replaced by the maximum of the absolute values in order to maximize the differences depending on the type of dynamical features being intensified.

Let us define a phase transition sequence as the sequence of characteristic exponents for a system $M$ running for longer runtimes as shown in Figure 14.

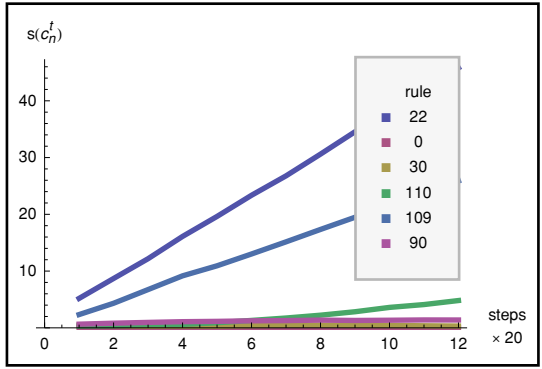

**Figure 14**. Phase characteristic exponent sequence $s(c_n^t(M))$.

The general rule for $t = 200$ and $n = 40$ is that if the characteristic exponent $c_n^t$ is greater than 1 for large enough values of $n$ and $t$, then $c_n^t$ has a phase transition. Otherwise it does not. Table 1 shows the calculation of the characteristic exponents of some elementary cellular automata.

| $S_c$ | $f(S_c)$ |
|---|---|
| {3.0, 5.2, 7.5, 9.9, 12., 15., 17., 20., 22., 24., ...} | $0.0916163 + 2.4603\,x$ |
| {2.6, 2.3, 3.6, 5.0, 6.7, 8.6, 9.6, 11., 12., 14., ...} | $-0.0956435 + 1.39588\,x$ |
| {1.5, 2.3, 3.2, 4.6, 5.8, 7.1, 8.7, 10., 11., 13., ...} | $-0.35579 + 1.28849\,x$ |
| {4.0, 6.3, 8.6, 11., 13., 15., 17., 20., 22., 24., ...} | $1.48149 + 2.30786\,x$ |
| {2.5, 3.3, 4.4, 5.6, 6.6, 7.3, 7.5, 8.0, 8.7, 9.1, ...} | $3.52132 + 0.492722\,x$ |
| {3.8, 4.3, 4.7, 5.0, 5.4, 5.6, 5.9, 6.4, 6.7, 7.2, ...} | $3.86794 + 0.296409\,x$ |
| {2.4, 3.7, 4.6, 5.4, 5.7, 6.0, 6.2, 6.4, 6.6, 6.7, ...} | $4.2508 + 0.184981\,x$ |
| {1.8, 1.8, 2.1, 2.8, 3.4, 3.8, 4.1, 4.4, 4.7, 4.8, ...} | $1.83839 + 0.270672\,x$ |
| {1.9, 3.0, 2.8, 3.3, 3.7, 4.0, 4.4, 4.7, 4.9, 5.0, ...} | $2.57937 + 0.207134\,x$ |
| {2.0, 3.1, 3.1, 3.7, 4.3, 4.6, 4.8, 5.1, 5.4, 5.4, ...} | $2.89698 + 0.218607\,x$ |
| {0.61, 0.45, 0.38, 0.39, 0.28, 0.30, 0.24, 0.28, 0.35, 0.43, ...} | $0.41144 - 0.00298547\,x$ |
| {0.35, 0.31, 0.40, 0.42, 0.56, 0.62, 0.72, 0.90, 1.2, 1.4, ...} | $-0.751501 + 0.268561\,x$ |
| {0.48, 0.41, 0.29, 0.37, 0.42, 0.42, 0.47, 0.51, 0.52, 0.55, ...} | $0.302027 + 0.0263495\,x$ |
| {0.087, 0.057, 0.038, 0.036, 0.027, 0.028, 0.024, 0.019, 0.017, 0.021, ...} | $0.0527182 - 0.0028416\,x$ |

**Table 1**.

### 5.1 Regression Analysis

Let $S_c = S(c_n^t)$ for a fixed $n$ and $t$. The line that better fits the growth of a sequence $S_c$ can be found by calculating the linear combination that minimizes the sum of the squares of the deviations of the elements. Let $f(S_c)$ denote the line that fits the sequence $S_c$ by finding the least-squares as shown in Figure 15.

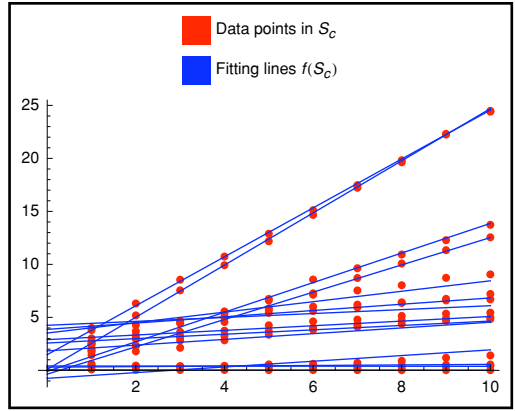

**Figure 15**.

The derivatives of a phase transition function are therefore stable indicators of the degree of the qualitative change in behavior of the systems. The larger the derivative, the larger the significance. Let $C$ denote the transition coefficient defined as $C = f'(S_c)$. Table 2 illustrates the calculated transition coefficients for a few elementary cellular automata rules.

### 5.1.1 Interesting Initial Conditions

After calculating the transition coefficient, we can calculate the first 10 most interesting initial conditions for elementary cellular automata with transition coefficients greater than 1. Those listed in Table 3 were calculated up to 600 steps in blocks of 50.

| ECA Rule (*r*) | C(ECA$_r$) |
|---|---|
| 22 | 2.5 |
| 151 | 2.3 |
| 109 | 1.4 |
| 73 | 1.3 |
| 133 | 0.49 |
| 183 | 0.30 |
| 54 | 0.27 |
| 110 | 0.27 |
| 97 | 0.22 |
| 41 | 0.21 |
| 147 | 0.18 |
| 45 | 0.026 |
| 1 | −0.0028 |
| 30 | −0.0030 |

**Table 2**. Phase transition coefficient.

| ECA Rule | Initial Configuration Number |
|---|---|
| 151 | {17, 18, 20, 22, 26, 34, 37, 41, 44, 46} |
| 22 | {8, 14, 17, 20, 23, 24, 26, 27, 28, 29} |
| 73 | {10, 12, 15, 16, 21, 24, 28, 30, 43, 45} |
| 109 | {2, 3, 11, 13, 20, 24, 26, 28, 32, 33} |
| 133 | {4, 6, 8, 10, 14, 16, 18, 19, 22, 25} |
| 94 | {10, 16, 20, 23, 24, 26, 28, 29, 30, 32} |

**Table 3**. Elementary cellular automata that are the most sensitive to initial configurations.

## 5.2 Phase Transition Classification

The coefficient *C* has positive values if the system is sensitive to the initial configurations. The larger the positive values, the more sensitive the system is to initial configurations. *C* has negative values or is close to 0 if it is highly homogeneous. Irregular behavior yields nonlinear growth, leading to a positive exponent.

For elementary cellular automata, it was found that $n = 40$ and $t = 200$ were runtime values large enough to detect and distinguish cellular automata having clear phase transitions. It is also the case that systems showing no quick phase transition have an asymptotic probability 1 of having a transition at a later time. In other words, a

$t = 200$    $n = 40$

system with a phase transition either has the transition very early in time or is unlikely to ever have one later, as can be theoretically predicted from an algorithmic theoretical argument. (A phase transition is undecidable and can be seen as a reachability problem equivalent to the halting problem, hence a system powerful enough to halt when reaching a phase transition, as calculated earlier, has an [effective] density zero [23].) The same transition coefficient can also be seen as a homogeneity measure. At the right granular level, randomness shares informational properties with trivial systems. Like a trivial system, a random system is incapable of transmitting or carrying out information. The characteristic exponent relates these two behaviors in an interesting way since the granularity of a random system for a runtime large enough is close to the dynamical state of being in a stable configuration according to this measure. One can see that rule 110 is better classified, certainly because it has some structure and is less homogeneous in time, unlike rule 30 and of course rule 1. Rule 30, like rule 1, changes its compressed output from one step to the other at a lower rate or not at all. While the top of the classification and the gap between them are more significant because they show a qualitative change in their evolution, the bottom is classified by its lack of changes. In other words, while rules like 22 and 151 exhibit more changes when starting from different initial classifications, rules such as 30 and 1 always look alike.

The clusters formed (a different cluster per row) for a few selected cellular automata rules starting from random initial conditions are shown in Figure 16.

The clusters shown in Figure 16 are clearly classifying this small selection of cellular automata by the presence of phase transitions (sudden structures). This is also a measure of homogeneity.

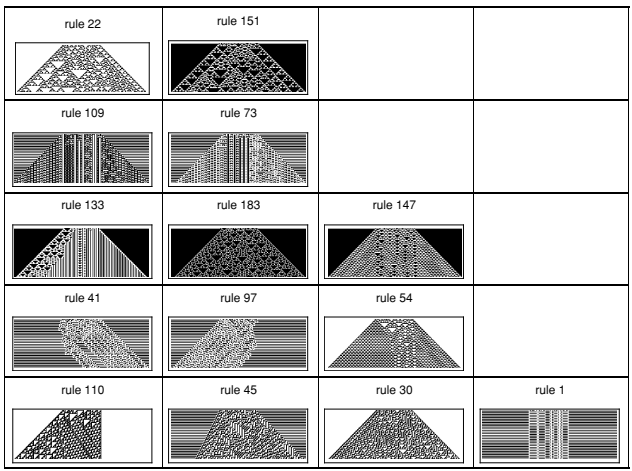

**Figure 16**. Clusters found using the phase transition coefficient over a sample of 14 rules.

A measure of homogeneity for classifying elementary automata according to their transition coefficients can be calculated. The top 24 and bottom 22 cellular automata up to 600 steps in blocks of 50 for the first 500 initial conditions followed by their transition coefficients sorted from larger to smaller values are shown in Figures 17 and 18. The complete table of transition coefficients is available at http://www.mathrix.org/experimentalAIT/.

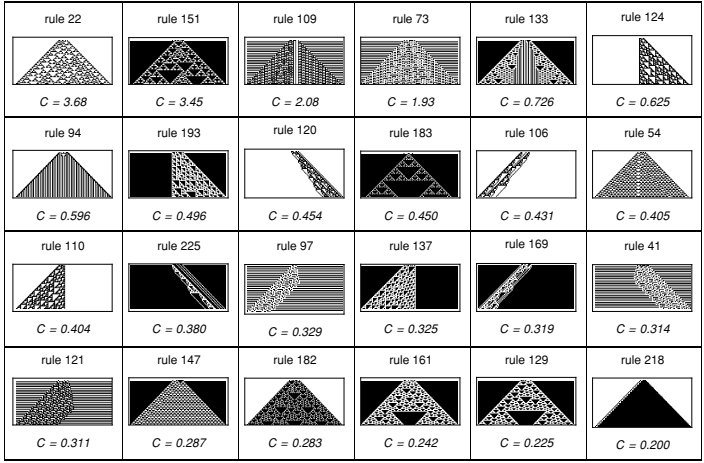

**Figure 17.** Phase transition coefficient classification (top 24) of elementary cellular automata

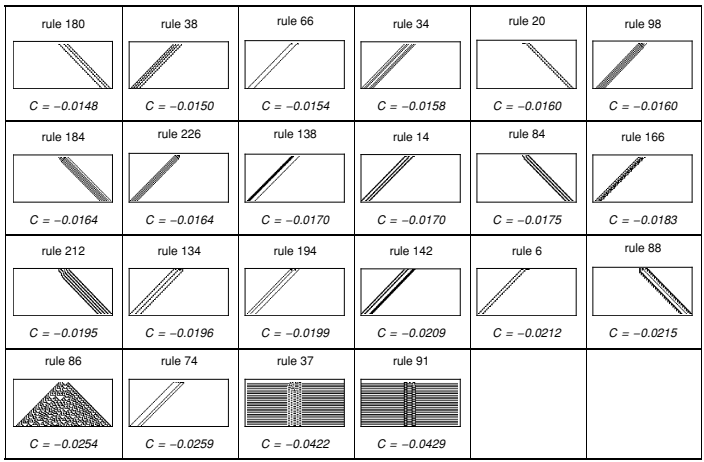

**Figure 18.** Phase transition coefficient classification (bottom 22) of elementary cellular automata

Since a random system would be expected to produce a homogeneous stream with no distinguishable patterns as incapable of carrying or transmitting any information, both the simplest and the random systems were classified together at the end of the figure.

### 5.2.1 Conjecture Relating Universality and the Phase Transition Coefficient

Based on this study, we conjecture that a system will be capable of universal computation if it has a large transition coefficient, at least larger than zero, say. The inverse, however, should not hold, because having a large transition coefficient by no means implies that the system will behave with the freedom required of a universal system if it is to emulate any possible computation (a case in point may be rule 22, which, despite having the largest transition coefficient, may not be versatile enough for universal computation). We base this conjecture on two facts:

1. The only known universal elementary cellular automata figure at the top of this classification, and so do candidates such as rule 54 which figures right next to rule 110.

2. Universality seems to imply that a system should be capable of being controlled by the inputs, which our classification suggests those at the bottom are not, as all of them look the same no matter what the input, and may not be capable of carrying information through the system toward the output.

The conjecture also seems to be in agreement with Wolfram's claim that rule 30 (as a class 3 elementary cellular automaton) may be, according to his Principle of Computational Equivalence (PCE), computationally universal. But it may turn out that it is too difficult (maybe impossible) to control in order to perform a computation because it behaves too randomly.

It is worth mentioning that this method is capable of capturing many of the subtle different behaviors a cellular automaton is capable of, which are heuristically captured by Wolfram's classes. The technique does not, however, disprove Wolfram's principle of irreducibility [2] because it is an a posteriori method. In other words, it is only by running the system that the method is capable of revealing the dynamical properties. This is no different from a manual inspection in that regard. However, it is of value that the method presented does identify a large range of qualitative properties without user intervention that other techniques (a priori techniques), including several analytical ones, generally seem to neglect.

## 6. Conclusion

We were able to clearly distinguish the different classes of behavior studied by Wolfram. By calculating the compressed lengths of the out-put of cellular automata using a general ompression algorithm we found two clearly distinguishable main clusters and, upon closer inspection, two others with clear gaps in between. That we found two main large clusters seems to support Wolfram's Principle of Computational Equivalence (PCE) [2], which suggests that there is no essential distinction between the classes of systems showing trivial and nested behavior and those showing random and complex behavior. We have also provided a compression-based framework for phase transition detection, and a method to calculate an exponent capable of identifying and measuring the significance of other dynamical properties, such as sensitivity to initial conditions, presence of structures, and homogeneity in space or regularity in time. We have also formulated a conjecture with regard to a possible connection between its transition coefficient and the ability of a system to reach computational universality. As can be seen from the experiments presented in this paper, the compression-based approach and the tools that have been proposed are highly effective for classifying, clustering, and detecting several dynamical properties of abstract systems. Moreover, the method does not depend on the system and can be applied to any abstract computing device or to data coming from any source whatsoever. It can also be used to calculate prior distributions and make predictions regarding the future evolution of a system.

## Acknowledgments

I would like to thank Jean-Paul Delahaye for his valuable suggestions, and Chiara Basile, Matthew Szudzik, Paul-Jean Letourneau, Jamie Williams, and Todd Rowland for their suggestions, stimulating discussions, and disseminating some of the ideas of this paper.